\documentclass[aps,prl,twocolumn,superscriptaddress]{revtex4}

\usepackage{graphicx}
\usepackage[amssymb]{SIunits}             	
\usepackage[usenames,dvipsnames]{color}   
\usepackage{amsmath}

\definecolor{orange}{rgb}{1,0.6,0}

\newcommand{\be}{\begin{equation}}
\newcommand{\ee}{\end{equation}}

\begin{document}

\title{Ultimate Turbulent Taylor-Couette Flow}


\author{Sander G. Huisman}  
\author{Dennis P.M. \surname{van Gils}}
\affiliation{Department of Applied Physics and J. M. Burgers Centre for Fluid Dynamics, University of Twente, P.O. Box 217, 7500 AE Enschede, The Netherlands}
\author{Siegfried Grossmann}
\affiliation{Department of Physics, Renthof 6, University of Marburg, D-35032 Marburg, Germany}
\author{Chao Sun}
\author{Detlef Lohse}
\affiliation{Department of Applied Physics and J. M. Burgers Centre for Fluid Dynamics, University of Twente, P.O. Box 217, 7500 AE Enschede, The Netherlands}

\date{\today}

\begin{abstract} 
The flow structure of strongly  turbulent Taylor-Couette flow with Reynolds numbers up to  $Re_i = 2 \cdot 10^6$ of the inner cylinder is experimentally examined with high-speed particle image velocimetry (PIV). The wind Reynolds numbers $Re_w$ of the turbulent Taylor-vortex flow is found to scale as $Re_w \propto Ta^{1/2}$, exactly as predicted \cite{gro11} for the ultimate turbulence regime, in which the boundary layers are turbulent. The dimensionless angular velocity flux has an effective scaling of $Nu_\omega \propto Ta^{0.38}$, also in correspondence with turbulence in the ultimate regime. The scaling of $Nu_\omega$ is confirmed by {\it local} angular velocity flux measurements extracted from high-speed PIV measurements: though the flux shows huge fluctuations, its spatial and temporal average nicely agrees with the result from the global torque measurements.
\end{abstract}

\maketitle

The Taylor-Couette (TC) system is one of the fundamental geometries conceived in order to test theories in fluid dynamics. Fluid is confined between two coaxial, differentially rotating cylinders. The system has been used to measure viscosity, study hydrodynamic instabilities, pattern formation, and the flow was found to have a very rich phase diagram \cite{and86}. In the fully turbulent regime, the focus up to now has been on {\it global} transport quantities \cite{lat92,lat92a,lew99,gil11,pao11}, which can be connected to the torque $\tau$, which is necessary to keep the inner cylinder rotating at constant angular velocity. In ref.\ \cite{eck07b} the analogy between the angular velocity flux in TC turbulence and the heat flux in Rayleigh-B\'enard (RB, see ref.\ \cite{ahl09}) flow was worked out, suggesting to express the former in terms of the Nusselt number, $Nu_\omega$, which in ref.\ \cite{gil11} was found to have an effective scaling  $Nu_\omega \propto Ta^{0.38}$ with the Taylor number (the analog to the Rayleigh number $Ra$ in RB flow). Such effective scaling $Nu \propto Ra^{0.38}$ characterizes the so-called ultimate scaling regime in RB flow \cite{cha97,cha01,ahl11a}. Following these papers, Grossmann and Lohse \cite{gro11} have interpreted this scaling as signature of turbulent boundary layers. They derived $Nu \propto Ra^{1/2} \times \hbox{log-corrections}$ (RB) and $Nu_\omega \propto Ta^{1/2} \times \hbox{log-corrections}$ (TC). The log-corrections imply the effective scaling law exponent of $0.38$. They also made a prediction for the accompanying scaling of the wind Reynolds number $Re_w$, namely 
\be 
Re_w \propto Ra^{1/2} ~ \hbox{and} ~
Re_w \propto Ta^{1/2} \label{gl}
\ee for RB and TC turbulence, respectively.
 Here the logarithmic corrections remarkably cancel out, in contrast to what Kraichnan had predicted \cite{kra62} earlier, namely
\be
Re_w \propto Ra^{1/2} (\log Ra)^{-1/2} 
~\hbox{or } ~
Re_w \propto Ta^{1/2} (\log Ta)^{-1/2},
\label{k62}
\ee
which leads to an effective scaling exponent of about 0.47 in the relevant turbulent regime. 
 In order to verify the interpretation of ref.\ \cite{gro11} and to check the prediction (\ref{gl}), {\it local} flow measurements are required to  extract the wind Reynolds number $Re_w$. However, what happens locally, inside the TC flow, has up to now only been studied for relatively low Reynolds numbers $Re < 10^5$, and has been restricted to flow profiles and single-point statistics \cite{wen33, col65a, pfi81, smi82, mul82, lor83, lat92, lew99, she01, lan04, abs08, rav10}. 

In this paper we supply local flow measurements from high-speed particle image velocimetry (PIV) at strongly turbulent TC flow. From these we will verify that indeed $Re_w \propto Ta^{1/2}$. In addition, from the PIV measurements we are able to also extract {\it local} angular velocity fluxes. These are found to strongly fluctuate in time, but when averaged azimuthally, radially, and in time, for the lower $Ta$ show a slight axial dependence, which we interpret as reminiscence of the turbulent Taylor vortices, and which nearly vanishes for the largest $Ta$ we achieve. 

The apparatus used for the experiments has an inner cylinder with a radius of $r_i = \unit{0.200}{\meter}$, a transparent outer cylinder with an inner-radius of $r_o = \unit{0.279}{\meter}$, resulting in a gap-width of $d= r_o-r_i = \unit{0.079}{\meter}$ and a radius ratio $\eta=r_i/r_o=0.716$. The height is $L=\unit{0.927}{\meter}$ implying an aspect ratio of $\Gamma=L/(r_o-r_i)=11.7$. More details regarding the experimental facility can be found in ref.\ \cite{gil11a}. Here we focus on the case of inner cylinder rotation and fixed outer cylinder. The local velocity is measured using PIV. We utilize the viewing ports in the top plate of the apparatus to look at the flow from the top. The flow is illuminated from the side using a pulsed Nd-YLF laser \cite{pivlaser}, creating a horizontal laser sheet. The working fluid (water) is seeded with \unit{20}{\micro \meter} polyamide seeding particles, and is recorded using a high speed camera \cite{photron}. The PIV system is operated in double-frame mode which allows us to have a $\Delta t$ far smaller than $1/f$, where $f$ is the frame rate.  The PIV measurements give us direct access to both the angular velocity $\omega (\theta, r,z, t) = u_\theta (\theta, r , z ,t)/r$ and the radial velocity $u_r (\theta , r, z, t)$, simultaneously.

\begin{figure}[ht!]
 \begin{center}
  \includegraphics{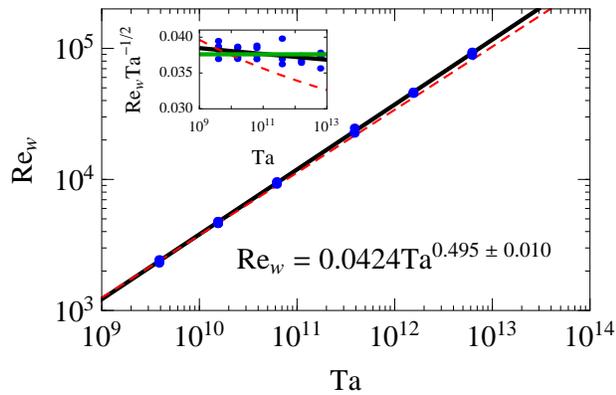}
  \caption{(color online) $Re_w$ vs. $Ta$.  The data from repeated experiments at mid-height are plotted as separate (blue) dots, showing the quality of the reproducibility and the statistical stationarity of the measurements. We have averaged azimuthally, over time, and in the bulk flow ($0.23\meter \le r \le 0.25\meter$). The straight line is the best fit $Re_w = 0.0424 Ta^{0.495 \pm 0.010}$ and the (red) dashed line is the Kraichnan prediction \cite{kra62} eq.\ (\ref{k62}). The inset shows the compensated plot $Re_w/Ta^{1/2}$ vs $Ta$. The horizontal (green) line is the prediction (\ref{gl}) of ref.\ \cite{gro11}.  
}
  \label{fig:RewvsTa}
 \end{center} 
\end{figure}

From the latter we extract the wind Reynolds number as $Re_w := u_{r}^{\text{std}} d/\nu$, where $u_r^{\text{std}}$ is the standard deviation of the radial velocity. In fig.\ \ref{fig:RewvsTa} $Re_w$ is shown as a function of the Taylor number
\begin{equation}
 Ta = \frac 14 \sigma (r_o-r_i)^2 (r_i+r_o)^2 (\omega_i - \omega_o)^2 / \nu^2  \label{deftaylor}.  
\end{equation}
In refs.\ \cite{eck07b,gil11} $Ta$ had been suggested as most appropriate independent variable of the TC system in order to work out the analogy with RB. Here $\sigma = \left(  (1+\eta) /(2 \sqrt{\eta} ) \right)^4$ can be interpreted as geometric ``Prandtl number'' \cite{eck07b}, $\omega_{i,o}$ is the angular velocity of the inner and outer cylinder, respectively, and $\nu$ is the kinematic viscosity. Note that $Ta \propto (\omega_i - \omega_o)^2$: while $Ra$ in RB convection is proportional to the temperature difference times the given gravity force, $Ta$  in TC flow is proportional to the angular velocity difference $\omega_i - \omega_o$ times the centrifugal force, which itself is also proportional to $\omega_i - \omega_o$, implying the square-dependence. Therefore, by definition, the two control parameters $Re_i$ (refering to the imposed azimuthal velocity) and $Ta$ are connected by $Re_i \sim Ta^{1/2}$, but such a trivial relation of course does not exist between the wind Reynolds number $Re_w$ and $Ta$ (which is a response of the systems and refers to the radial velocity). 
 
Fig.\ \ref{fig:RewvsTa} reveals a clear scaling of the wind Reynolds number with the Taylor number, namely $Re_w \propto Ta^{0.495\pm0.010}$, which is consistent with the prediction \cite{gro11} $Re_w \propto Ta^{1/2}$ for the ultimate TC regime, but inconsistent with Kraichnan's earlier prediction (\ref{k62}) of a scaling exponent $1/2$ {\it with} logarithmic corrections \cite{kra62}.
For comparision, we included this  relation into fig.\ \ref{fig:RewvsTa}, which clearly is inconsistent with the experimental data. We stress that the cancellation of the log-correction for $Re_w$ as suggested in \cite{gro11} is highly non-trivial and that in RB flow in the  non-ultimate regimes the wind Reynolds number scales as $Re_w \sim Ra^{0.44}$ \cite{qiu01b}, pronouncedly different than the 1/2 exponent we find here in the ultimate regime. Only very recently the wind Reynolds number scaling in ultimate RB flow could be measured, also finding $Re_w \sim Ra^{1/2}$ \cite{he11} as predicted in ref.\ \cite{gro11}.

\begin{figure*}[ht!]
 \begin{center}
    \includegraphics{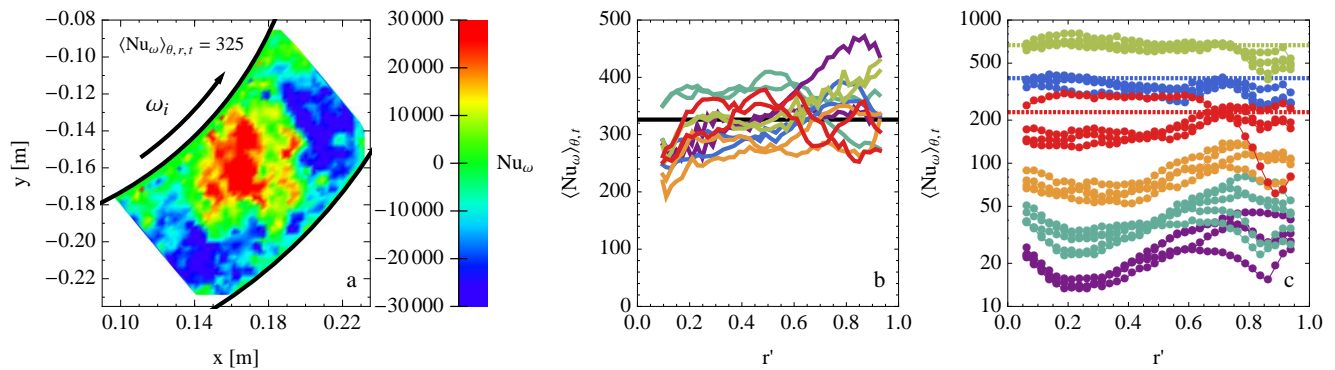}
  \caption{(color online) 
(a) Snapshot of the instantaneous convective angular velocity flux, measured at $z=L/2$, for $Ta=1.5 \cdot 10^{12}$. The $(r,\theta)$-plane and time averaged flux is found to be equal to $\left \langle Nu_\omega \right \rangle_{\theta, r, t} = 325$. A corresponding movie is available as supplementary material. 
(b) Local normalized convective $\omega$-flux as functions of $r' = (r - r_i)/(r_o - r_i)$ for 6 heights varying between $0.5 \leq z/L \leq 0.73$, for $Ta=1.5 \cdot 10^{12}$. The black solid line is the average of the 12 experiments ($\langle Nu_\omega \rangle_{\theta, r, z, t}$), which is very close to the expected value $326 \pm 6$ from global torque measurements \cite{gil11}. 
(c) Local normalized convective angular velocity flux vs. radial position $r'$ for various rotation rates, measured at 
$z=L/2$. From bottom to top we have repeated experiments for $Ta=3.8\cdot 10^9$, $1.5\cdot 10^{10}$, $6.2\cdot 10^{10}$, $3.8\cdot 10^{11}$, $1.5\cdot 10^{12}$, and $6.2\cdot 10^{12}$. The dashed lines for the three highest $Ta$ represent the $Nu_\omega$ value derived from global torque measurements \cite{gil11}. 
}
  \label{fig:localtransport}
 \end{center} 
\end{figure*}

Next, as the PIV measurements give us both the angular velocity $\omega(\theta, r, z, t)$ and the radial velocity $u_r(\theta, r, z, t)$, we can directly calculate the (total) angular velocity flux (convective $+$ molecular)  
\begin{align}
 J^\omega (\theta, r,z, t) & := r^3 \left( u_r \omega - \nu \partial_r \omega  \right),  \label{jomega}
\end{align}
which is made dimensionless with its value for the laminar infinite aspect ratio case, $J^\omega_\text{lam} = 2 \nu r_i^2 r_o^2 (\omega_i - \omega_o) / (r_o^2 - r_i^2)$, giving \cite{eck07b} the local ``Nusselt number''
\begin{align}
 Nu_\omega  (\theta , r, z, t) &= J^\omega (\theta , r, z, t) / J^\omega_\text{lam}. 
\end{align}
Indeed, as shown in ref.\ \cite{eck07b}, the  angular velocity is the relevant quantity transported from the inner to the outer cylinder, as its flux (\ref{jomega}) is radially conserved,  once it is averaged azimuthally, axially, and over time, $\frac{d}{dr} \left \langle J^\omega (\theta, r, z, t) \right \rangle_{\theta, z, t} = 0$. In the turbulent regime the convective term is the major contributor to the flux in the bulk \cite{ldaunpublished}.

In fig.\ \ref{fig:localtransport}a we show a snapshot of $Nu(\theta ,r)$ at mid-height $z=L/2$ for $Ta = 1.5 \cdot 10^{12}$. The quantity shows {\it huge} fluctuations, ranging from $+10^5$ to $-10^5$ and beyond, whereas the average $\left\langle Nu_\omega (\theta , r, t)\right \rangle_{\theta , r, t} = 325$ is very close to the value $Nu_\omega^{glob}= 326 \pm 6$ obtained from global torque measurements \cite{gil11}. 
The local flux can thus be  more than $\pm 300$ times as large as the mean flux. Large fluctuations have also been reported for the local heat-flux in RB flow \cite{sha03}, but in that case the largest fluctuations were only 25 times larger than the mean flux. 

After azimuthal and time averages, $\left\langle Nu_\omega (\theta , r, t)\right\rangle_{\theta,t}$, the fluctuations nearly vanish, see fig.\ \ref{fig:localtransport}b (revealing some radial and height dependence for fixed $Ta = 1.5 \cdot 10^{12}$, presumably reminisent of the Taylor vortices) and fig.\  \ref{fig:localtransport}c, where we show the local angular velocity flux $r'$-profiles for rotation rates from $\omega_i/(2\pi) = \unit{0.5}{\hertz}$ to $\unit{20}{\hertz}$, corresponding to $Ta=3.8\cdot 10^9$ to $6.2\cdot 10^{12}$. Each profile is based on azimuthal averaging, radial binning, and averaging over 3200 frames (corresponding to 25.6 rotations for the three lowest rotation rates, and 32, 64, and 128 rotations for the fastest rotations rates). For each rotation rate repeated experiments have been performed  and the profiles are reproducible. Only in one case the turbulent Taylor vortex flow seems to be in a different state(s). From fig.\ \ref{fig:localtransport}c  we conclude that the spread in the repeated experiments decreases with increasing $Ta$, for which the Taylor vortex structure  will be more and more washed out. In addition, for increasing $Ta$, not only do we measure during more revolutions, but also the transverse velocity increases, both improving the statistics. 
The dashed lines in fig.\ \ref{fig:localtransport}c correspond to the measured global transport for the three highest rotation rates; these values were obtained from the torque measurements \cite{gil11} and show already good agreement with our local measurements. 

\begin{figure}[ht!]
 \begin{center}
  \includegraphics{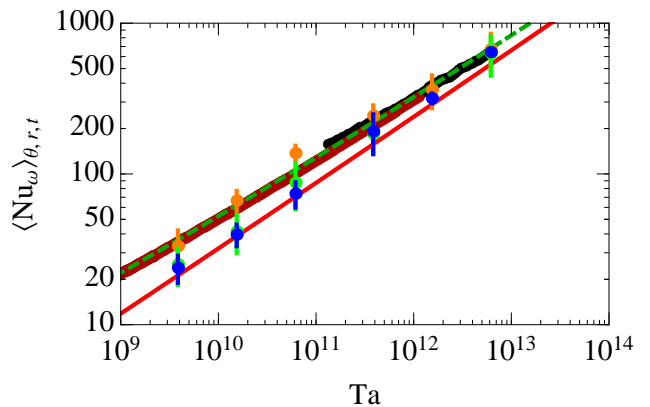}
  \caption{(color online) 
Local convective angular velocity flux as a function of Taylor number. The blue dots are results obtained from PIV measurements and show a scaling of $Nu_\omega \propto Ta^{0.45\pm0.04}$. The green and orange dots are repeated measurements at a height of $z=L/2+d/2$ and $z=L/2+ d$, respectively. The black data points are obtained from global torque measurements and show a scaling that is less steep: $Nu_\omega \propto Ta^{0.38}$. The dashed green line is obtained by matching two log-layers \cite{lat92a}, and has a slope of $0.37$ at $Ta=10^9$, and $0.41$ at $Ta=10^{13}$. The red line is from the turbulent boundary layer theory of ref.\ \cite{gro11}. It has a slope of $0.43$ around $Ta=10^9$ and $0.44$ around $Ta=10^{13}$. Dark red data points are obtained by means of global torque measurements \cite{lew99}.}
  \label{fig:NuvsTa}
 \end{center} 
\end{figure}

An additional {\it axial} average is necessary to obtain the exact relation between $Nu_\omega$ and the global torque $\tau$  required to drive the inner cylinder at constant velocity \cite{eck07b}, 
\begin{equation}
\tau = 2\pi L \rho J_{lam}^\omega  \left\langle Nu_\omega \right\rangle_{\theta, z, t} . 
\label{glo-lo-link}
\end{equation}
It is the lack of sufficient axial averaging, which accounts for the small deviations between $\left\langle Nu_\omega (\theta, z, t) \right\rangle_{\theta, r, t}$ and $Nu_\omega^{glob}$. Indeed, due to the Taylor-vortex structure of the TC flow one would expect some axial dependence of  $\left\langle Nu_\omega (\theta , r, z, t) \right\rangle_{\theta, r, t}$, which should become weaker with increasing degree of turbulence and thus increasing $Ta$, just as fig.\ \ref{fig:localtransport}c suggests. This picture is confirmed in figure \ref{fig:NuvsTa}. Here we present local measurements of the convective angular velocity flux for varying rotation rates, resulting in a Taylor number range of $3.8\cdot10^9$ -- $6.2 \cdot 10^{12}$. For each Taylor number we performed multiple experiments and measured the $Nu_\omega$ transport at mid-height. The blue points are results obtained from PIV measurements at mid-height, where the length of the bars indicate the error obtained from the repeated experiments. The green and orange points are repeated measurements at $z=L/2+d/2$ and $z=L/2+d$, respectively. An effective scaling $Nu_\omega \propto Ta^{0.45\pm0.04}$  is revealed for the blue data points, while a scaling of $Nu_\omega \propto Ta^{0.39\pm0.08}$ is revealed for the orange data points. 

\begin{figure}[!t]
 \begin{center}
  \includegraphics{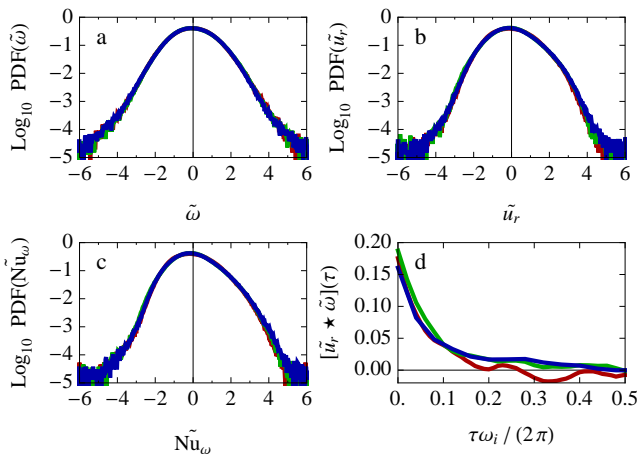}
  \caption{(color online) Results of three experiment with varying rotation rate resulting in $Ta=3.8 \cdot 10^{11}$, $1.5 \cdot 10^{12}$, and $6.2 \cdot 10^{12}$, colored in red, green, and blue, respectively. All the data shown is averaged over the region $\unit{0.23}{\meter} \leq r \leq \unit{0.25}{\meter}$, and measured at mid-height. All quantities with tildes are standardized (shifted and scaled such as to have zero mean and unit variance). 
(a) PDF of the standardized angular velocity. 
(b)  PDF of the standardized radial velocity. 
(c) Standardized normalized local convective angular velocity flux PDF.
 (d)  Cross-correlation coefficient of the angular velocity and the radial velocity, the dimensionless decaying time (in number of rotations) is found to be 0.07. The corresponding length scale can be found by multiplying this number with the circumference of the inner cylinder giving $\delta = \unit{88}{\milli \meter}$, which is of the same order of magnitude as the gap width $d  = \unit{80}{\milli \meter}$.}
  \label{fig:PDF_corr}
 \end{center} 
\end{figure}

It is remarkable how the flow provides angular velocity transport from the inner to the outer cylinder, in spite of the fluctuative nature, which are seen in figure \ref{fig:localtransport}a. In fig.\ \ref{fig:PDF_corr} we provide a statistical analysis of these fluctuations: While the probability distribution functions (PDFs) of the angular velocity (fig.\ \ref{fig:PDF_corr}a) and the radial velocity (fig.\ \ref{fig:PDF_corr}b) are nearly symmetric, the PDF of their product  $r^3 u_r \omega \propto Nu_\omega $ (fig.\ \ref{fig:PDF_corr}c) is clearly positively skewed. Indeed, the cross-correlation coefficient of $u_r$ and $\omega$ (fig.\ \ref{fig:PDF_corr}d) is relatively large.

We note that thanks to the PIV measurements of the full velocity field, the extraction of the local angular velocity flux $Nu_\omega (\theta ,r, z, t) \propto \omega u_r$ is easier in TC as compared to the analog temperature flux $Nu(\vec x , t) \propto T u_z$ in RB flow: in order to obtain  this latter quantity locally, one has to measure the temperature $T$ and the velocity simultaneously. Because a high-precision field-measurement of the temperature is presently not possible and thus not available, the best one can do for RB flow is to measure $Nu (\vec x ,t)$ point by point \cite{sha03,sha08} or use an instrumented tracer \cite{she07}. 


In conclusion, from high-speed PIV measurements we have found the wind Reynolds number in strongly turbulent TC flow to scale as  $Re_w \propto Ta^{0.495\pm0.010}$,  in accordance with the theory of ref.\ \cite{gro11} and in conflict with Kraichnan's \cite{kra62} prediction (\ref{k62}). In addition, we extracted  the local  angular velocity flux and found that $Nu_\omega \propto Ta^{\gamma}$ with $\gamma \sim 0.39-0.45$ depending on the axial position and consistent with earlier global torque measurements \cite{gil11,pao11}. For increasing $Ta$, a small axial dependence of $Nu_\omega$ is fading away, reflecting the decreasing importance of the Taylor vortices. The next step will be to provide full velocity and angular velocity profile measurements, including those in the boundary layers, and  to extend the present measurements to the counter-rotating case and other radii ratios $\eta$, in order to further theoretically understand the local flow organization and the interplay between bulk and boundary layers in turbulent TC flow. A further highly interesting support for the presented idea of the close correspondence between the TC angular velocity transport in the studied $Ta$-range with the ultimate range of RB thermal convection is to identify the onset of this ultimate range when increasing $Ta$; here we expect a change of the $Nu_{\omega}$ scaling exponent and also a transitional change in the widths and profiles of the BLs.   

\begin{acknowledgments}
This study was financially supported by the Technology Foundation STW of The Netherlands.
\end{acknowledgments}


\end{document}